**On the reproducibility of enzyme reactions and kinetic modelling**

Gudrun Gygli

Institute for Biological Interfaces (IBG 1), Karlsruhe Institute of Technology (KIT), 76344 Eggenstein-Leopoldshafen, Germany;

orcid.org/0000-0001-9119-1778; Email: gudrun.gygli@kit.edu

**Abstract**

Enzyme reactions are highly dependent on reaction conditions. To ensure reproducibility of enzyme reaction parameters, experiments need to be carefully designed and kinetic modelling meticulously executed. Furthermore, to enable the judgement of the quality of enzyme reaction parameters, the experimental conditions, the modelling process as well as the raw data need to be reported comprehensively. By taking these steps, enzyme reaction parameters can be open and FAIR (findable, accessible, interoperable, re-usable) as well as repeatable, replicable and reproducible. This review discusses these issues and provides a practical guide to designing initial rate experiments for the determination of enzyme reaction parameters and gives an open, FAIR and re-editable example of the kinetic modelling of an enzyme reaction. Both the guide and example are scripted with Python in Jupyter Notebooks and are publicly available (https://fairdomhub.org/investigations/483). Finally, the prerequisites of automated data analysis and machine learning algorithms are briefly discussed to provide further motivation for the comprehensive, open and FAIR reporting of enzyme reaction parameters.

*Motivation*

Enzymes are the catalysts enabling life, accelerating chemical reactions in organisms so they can metabolize nutrients, grow, and pass on their DNA. Enzyme reactions are also used in different industries (Gygli and Berkel, 2015; Bell *et al.*, 2021; Wu *et al.*, 2021; Žnidaršič-Plazl, 2021), for example lactases to produce lactose-free dairy (Dekker, Koenders and Bruins, 2019), proteases in cleaning products (Olsen and Falholt, 1998), and enzymes used to produce the cholesterol lowering drug Lipitor®(Ma *et al.*, 2010). Enzyme reactions are highly dependent on the experimental conditions and on the state of the enzyme (Yang, 2009; van Schie *et al.*, 2021). The complex interplay of these factors can change enzyme activities in unexpected ways, presenting a challenge to our understanding of enzyme reactions and molecular mechanisms (Fitzpatrick and Klibanov, 1991; Klibanov, 2001; Bauduin *et al.*, 2004). Enzyme reactions can be studied using different experimental approaches, namely initial rate, progress curve (Stroberg and Schnell, 2016), transient kinetics and relaxation experiments (Cornish-Bowden, 2012). By fitting appropriate models, the underlying molecular mechanisms and the associated kinetic and thermodynamic parameters (enzyme reaction parameters, ERPs) can be determined. Examples of parameters relevant for enzyme reactions are the turnover number, $k_{cat}$, the Michaelis(-Menten) constant (Johnson and Goody, 2011), $K_m$, but also binding parameters like the dissociation constant, $K_d$, and the Gibbs free energy of binding, $\Delta G^0$, which become especially relevant for multi-substrate enzyme reactions.

This review discusses how reproducibility can be ensured for enzyme reactions and reaction modelling with initial rate experiments. Firstly, the current situation will be described with a focus on why many enzyme reactions are not reproducible and why the quality of enzyme reaction modelling can often not be assessed. Secondly, practical considerations to obtain high-quality ERPs will be given. Thirdly, an illustrative example of ERPs will be shown. I conclude with an outlook on what can be achieved with reproducible, high-quality data, and arguments for why current shortcomings are detrimental to the application of sophisticated data analysis tools such as machine learning (ML).

*The current situation*

*Why are many ERPs not reproducible?* Science is reportedly in a "reproducibility crisis" (Baker, 2016), and reproducibility problems have also been reported for ERPs (Wittig *et al.*, 2014; Halling *et al.*, 2018). It appears that ERPs are reported without all experimental conditions and/or details on the reaction modelling process. Arguably, in this discussion a distinction between repeatability, replicability and reproducibility (**Figure 1,** (Plant *et al.*, 2014; McArthur, 2019)) should be made. For enzyme reactions and enzyme kinetics data, this distinction means answering the following three questions: Can the experimental and modelling steps used to determine ERPs be *repeated* by scientists in the same lab, using the same devices and software, and give comparable results? Can the experimental and modelling steps described be *replicated* by scientists in another lab, using the same devices and software, and give comparable results? Can the ERPs obtained from experimental and modelling steps be *reproduced* by scientists using different devices and software in another lab? Incompletely reported experimental conditions hinder all three of these levels. Likewise, accuracy, precision and robustness of measurements (Plant *et al.*, 2014) should be considered when comparing the reported data and judging its repeatability, replicability and reproducibility. Note that the focus must not only be on the experimental steps taken, but also on the reaction modelling steps. These computational steps are a major part of any experiment studying an enzyme reaction, and therefore must also be reported in sufficient detail so that they can be repeated, replicated and reproduced. In the following, the impact of the reaction conditions on the enzyme reaction and the reaction modelling process will be discussed.

*The influence of the reaction conditions*

The effect of the reaction conditions and the enzyme state on enzyme reactions can be closely linked and influence the reaction mechanism. The pH can change the enzyme mechanism or influence the redox potential of the cofactor (Rungsrisuriyachai and Gadda, 2009; Vogt *et al.*, 2014). Also consider that the pH of reaction buffers is strongly affected by the presence of high concentrations of ions.(Yang *et al.*, 2010) Therefore, the composition and ionic strength of the reaction buffer can have pleiotropic effects on protein and enzyme activities (Bauduin *et al.*, 2004, 2006; Žoldák, Sprinzl and Sedlák, 2004; Broering and Bommarius, 2005; Yao *et al.*, 2021). These "Hofmeister effects" are well known and described since the 1880s, yet their mechanisms remain elusive to this day(Kunz, Henle and Ninham, 2004; Kunz, Lo Nostro and Ninham, 2004). Additionally, organic (co-)solvents can affect enzyme activity and selectivity in surprising ways (Fitzpatrick and Klibanov, 1991; Klibanov, 2001), making "solvent engineering" of non-aqueous reaction media a prominent field of study(Bommarius and Paye, 2013; Sheldon and Woodley, 2018; van Schie *et al.*, 2021). Macromolecular crowding (Ellis, 2001), caused by the presence of high concentrations of macromolecules in cells is absent in most *in* vitro assays, but is known to also affect enzyme activities (Ma and Nussinov, 2013; Poggi and Slade, 2015). Finally, many enzymes denature if heated above a certain temperature (Robinson, 2015), yet a growing number of enzymes are known to be active under extreme conditions (Littlechild, 2015). Enzyme reactions also strongly depend on the state of the enzyme, i.e. its purity and the "homogeneity of the enzyme sample" (HES). The HES describes if the enzyme is monomeric, dimeric, aggregated or agglomerated (Association of Resources for Biophysical Research in Europe and Molecular Biophysics in Europe, no date).

*The reaction modelling process*

ERPs are very frequently determined using initial rate experiments. $k_{cat}$ is a measure of how many molecules of substrate an enzyme can convert per second, and $K_m = \frac{k_{cat}+k_{off}}{k_{on}}$ is the substrate concentration at which half the turnover number is reached. $k_{off}$ and $k_{on}$ are the average rates at which enzyme-substrate complexes dissociate and associate, respectively. $K_m$ is often interpreted as the binding affinity of the substrate, $K_d = \frac{k_{off}}{k_{on}}$. However, this assumption can be misleading. Leonor Michaelis and Maud Menten assumed that $k_{off}$, is much larger than $k_{cat}$ with their rapid equilibrium approximation. Yet, at the same time as Michaelis and Menten worked on their equation, Donald van Slyke and Glenn Cullen developed their own equation to describe enzyme kinetics (Van Slyke and Cullen, 1914). Van Slyke and Cullen assumed that product formation occurs much more rapidly than substrate dissociation, meaning that $k_{cat}$ is much larger than $k_{off}$ (Van Slyke and Cullen, 1914). This distinction can become relevant if a reaction involves multiple substrates, such as a cofactor-dependent reaction, and $K_d$ is found to be much smaller than $K_m$ (Ott *et al.*, 2021). Contributions by other authors to modern enzyme kinetics are discussed elsewhere (Cornish-Bowden, 2013), as is the fascinating life of Maud Menten (Skloot, 2000).

Not all enzymes follow Michaelis-Menten-kinetics, and enzymes can have different reaction mechanisms depending on the substrates and/or reaction conditions (Carunchio, Girelli and Messina, 1999; Rungsrisuriyachai and Gadda, 2009; Freiburger *et al.*, 2014; Vogt *et al.*, 2014; Romero *et al.*, 2018; Ott *et al.*, 2021). Many different kinetic models have been formulated to represent these different reaction mechanisms, such as models for different modes of inhibition, allosteric activation or ping-pong mechanisms (Cleland, 1967; Srinivasan, 2021). Therefore, any knowledge on the enzyme reaction mechanism must be matched with the biochemical implications of the model and it is not always simply a matter of picking a kinetic model that gives a good fit. Especially for multi-substrate reactions, additional information on the reaction mechanism should be considered and included. For example, the binding parameters of the reaction components can assist the fitting process of inhibition models by including them as the binding constant of the inhibitor, $K_i$, and ensure that the model matches the enzyme reaction mechanism (Ott *et al.*, 2021). Furthermore, information on enzyme inactivation, as obtained from a Selwyn test, can be vital to design further experiments that allow the fit with a representative model (Schnell and Hanson, 2007). Such "atypical" reaction mechanisms are becoming more and more important when modelling ERPs in metabolic pathways (Vasic-Racki, Kragl and Liese, 2003; Cornish-Bowden and Cárdenas, 2010) or ERPs of multi-protein complexes in drug discovery (Atkins, 2005; Srinivasan, 2021).

*How to…*

*How to experimentally obtain high-quality data on an enzyme reaction*

The study of enzyme reactions and enzyme mechanisms is an interdisciplinary field, and each experiment presents its own specific challenges to reproducibility and data quality. In many cases, the enzyme is produced through heterologous expression in a model organism. Therefore, the quality of the enzyme sample is the first issue that needs to be addressed. Guidelines have been formulated for quality control of protein samples (Association of Resources for Biophysical Research in Europe and Molecular Biophysics in Europe, no date), and two items will briefly be summarized here. Firstly, the purity of an enzyme sample is routinely determined using sodium dodecyl sulphate–polyacrylamide gel electrophoresis (SDS-PAGE) (Laemmli, 1970). With this method, proteins are denatured and separated based on size, thus allowing the identification of contaminating proteins. Secondly, the HES can be determined using analytical size exclusion chromatography (SEC). With this method, the components of a native enzyme sample are separated by size. Analytical SEC is used to distinguish between monomers, dimers and other oligomers of a purified protein and typically is low throughput. With Dynamic Light Scattering (DLS) (Lorber *et al.*, 2012), higher throughput can be reached in the determination of HES, at the cost of not being able to distinguish monomers from dimers and other oligomers. Reporting both the purity of the enzyme sample and its HES provides important information on the state of the enzyme during the experiment. HES can indicate if inactive enzyme in present in the sample, which can lead to overestimation when calculating the turnover number, $k_{cat}$. The HES therefore provides crucial insight if the activity of an enzyme changes unexpectedly between measurements, for example between different expression and purification batches or enzyme used again after long storage. Determination of the HES should therefore be routinely included in the experimental workflow.

Once the quality of the protein sample is established, the correct design of experiments to determine ERPs and their comprehensive reporting becomes relevant (Wittig *et al.*, 2014; Halling *et al.*, 2018). Guidelines concerning this issue have been created (*STRENDA Guidelines*, no date; Murphy, Gilmour and Crabbe, 2002; Scopes, 2002; Lorsch, 2014) and some items will be summarized below. Initial rate experiments are often performed using UV-Vis spectrophotometry by following the consumption of substrate or the formation of product if either of the two is photometrically active. However, the upper and lower detection limit need to be considered. If the concentration of the spectrophotometrically active molecule is above the upper detection limit, but at the same time below or close to $K_m$, the enzyme is not saturated with substrate. Therefore, the thus obtained ERP will be incorrect and an alternative method should be used to determine ERPs, for example isothermal titration calorimetry (ITC) (Ott *et al.*, 2021). Also, the extinction coefficient used to calculate the concentration of the spectrophotometrically active molecule can depend on the composition and pH of the reaction buffer (Yang *et al.*, 2010). Therefore, the extinction coefficient should be experimentally determined for new reaction buffers or solvents and must always be reported.

In initial rate experiments, reaction rates at different substrate concentrations, $[S_0]$, are measured. These reaction rates are considered to be "initial reaction rates" if enzyme concentrations, $[E_0]$, are sufficiently low so that $[S_0] \gg$ product concentration, $[P]$, and $[S_0] \gg [E_0]$. Under these conditions, the measured reaction rate is not affected by product accumulation or substrate depletion. $[E_0]$ is typically in low nM amounts. Initial reaction rates are obtained from the slope of a linear model fitted to the raw data (**Figure 2A**). Linear fits should be fitted for at least ten datapoints, and a good quality fit should be obtained, e.g. $r^2$ <0.9. If such a fit is not possible (**Figure 2B**), $[E_0]$, the spacing between measurements or the duration of the measurements should be adjusted to obtain more datapoints (**Figure 2C**).

These initial rates, measured at different [$S_0$], can then be plotted as a function of [$S_0$] in a Michaelis-Menten plot and fitted with the Michaelis-Menten equation to model the enzyme reaction. The choice of the different [$S_0$] is crucial to the quality of the ERPs, independent of the model chosen to fit the data. If possible, the lowest [$S_0$] should be at least 10x lower than $K_m$ and the highest [$S_0$] at least 10x higher than $K_m$, with six concentrations chosen below and four above $K_m$ (Murphy, Gilmour and Crabbe, 2002; Lorsch, 2014). This is recommended because the reaction rates change most at [$S_0$] below $K_m$. It can also be useful to more narrowly space [$S_0$] below or around $K_m$ and to more widely space [$S_0$] above $K_m$ (**Figure 3**).

If no literature data is available for the enzyme and the reaction of interest, an initial rough estimate of the kinetic parameters can be obtained through a "zero-round experiment" (**Figure 3A**). Widely spaced [$S_0$] can be used, for example five measurements with [$S_0$] spanning from low μM to high mM. Negative controls and control experiments should be included already at this stage to establish the stability of the enzyme and substrate for the duration of the experiment and eliminate any possible influence of other reaction components on the measured rate. From this zero-round experiment, a very rough estimate of the $K_m$ can be obtained (**Figure 3A**). Based on this estimate of $K_m$, a "first-round experiment" can be designed where a total of ten [$S_0$] are chosen in agreement with the considerations discussed above (**Figures 3B and 3C**). Again, this "first-round experiment" should also include negative controls and control experiments. Finally, a "gold-round experiment" can be designed with multiple replicates of all [$S_0$] and controls, e.g. using a 96-well microtiter plates (**Figure 3D**). Note that accuracy and reproducibility of data obtained in such experiments can be reduced (Grosch *et al.*, 2017), for example due to spatial and temporal temperature profiles in commercial microtiter plate readers (Grosch *et al.*, 2016).

In conclusion, to assess the quality of the reaction modelling process, and the ERPs $k_{cat}$ and $K_m$, two plots are needed. Firstly, the linear fits of the raw data used to calculate the initial reaction rates (**Figure 2**), and secondly the fits of the initial reaction rates with the Michaelis-Menten model (**Figure 3**).

*Additional experiments*

A "Selwyn test" (Selwyn, 1965) can provide insight into enzyme inactivation and can easily be included in a "gold-round experiment". To perform a Selwyn test, [$S_0$] is kept constant close to $V_{max}$ and three to five different [$E_0$] are used. [P] is then plotted against [$E_0$]*time. If the enzyme is not inactivated during the experiment, all points for the different [$E_0$] fall on the same curve and the Selwyn test is passed (Selwyn, 1965; Baici, 2015). A failed Selwyn test distinguishes between enzyme inactivation, where the curves do not overlap, and product inhibition, where the curves do not overlap and run parallel (Baici, 2015). Disentangling product inhibition from enzyme inactivation can be notoriously difficult with this approach because activities in the absence of product cannot be obtained.

ITC can be used to disentangle inhibition from enzyme inactivation (Ott *et al.*, 2021), and measure the kinetics of enzyme inhibition (Di Trani, Moitessier and Mittermaier, 2017). ITC is a highly sensitive, label-free analysis method that measures the heat discharged or consumed along a biomolecular reaction or interaction (Freire, Mayorga and Straume, 1990). ITC is still mostly used to measure thermodynamic parameters of biomolecular interactions, such as the binding parameters, $K_d$, $\Delta H_{binding}$, $\Delta G°$, $-T\Delta S$, and the stoichiometry of binding (Freyer and Lewis, 2008; Falconer, 2016; Roy *et al.*, 2020). Yet, ITC is used more and more to measure ERPs (Freyer and Lewis, 2008; Di Trani, Moitessier and Mittermaier, 2017; Zambelli, 2019). Three different types of ITC experiments can be performed: the "single-injection method", the "recurrent single-injection method" and the "multiple injection method" (Di Trani, Moitessier and Mittermaier, 2017). With the "single-injection method", $K_m$ and $k_{cat}$ can be determined in one measurement. In a variation of this experiment, the "recurrent single-injection method", substrate is injected repeatedly to study the effect of product accumulation, i.e. product inhibition or activation. The "multiple injection method" mimics an initial rate experiment. With this third method, substrate is titrated to the enzyme, achieving a stepwise increase in [S], and from each step, a reaction rate is obtained, enabling the calculation of $K_m$. Thus, the same information as from the ten measurements in an initial rate experiment can be obtained from only one measurement. While the throughput of ITC is lower than in assays based on 96-well microtiter plates, it provides the possibility to study a reaction in much more detail and also disentangle the impact of an amino acid on the reaction or binding process (Chen *et al.*, 2017).

ITC is especially worthwhile if multi-substrate enzymes are studied because it allows the determination of binding parameters of all the reaction components. These binding parameters can assist in the reaction modelling process to match knowledge on the enzyme reaction mechanism with the biochemical implications of the kinetic model of such complex reactions (Ott *et al.*, 2021). In conclusion, ITC experiments can provide a wealth of information on an enzyme reaction that is otherwise inaccessible.

*How to model ERPs*

Different software exists to obtain ERPs using different kinetic models. Most software requires manual intervention to fit a model. However, manual steps are problematic for two reasons. Firstly, they typically are more time consuming and prevent the upscaling of the analysis. A manual analysis is manageable for a handful of experiments, yet becomes limiting if hundreds of experiments need to be analyzed. Secondly, every manual step is a potential source of variation or error, making documentation of the analysis challenging, thus hindering repeatability, replicability and reproducibility. However, a scientist can write her own program(s), using for example Python and Jupyter Notebooks, to automate the analysis process and provide a clear and reusable documentation of exactly how she obtained her results. The continuing digitalization of science, growing interdisciplinarity(Islam and Wells, 2021) and interactions with large volumes of data (Carey and Papin, 2018) make it abundantly clear that every STEM (Science, Technology, Engineering, and Mathematics) scientist should know how to program (McDonald *et al.*, 2022). While it may initially seem daunting to learn a programming language like Python, the power and freedom it grants the budding programmer cannot not be overstated (Ayer, Miguez and Toby, 2014). Note that any automated data analysis is simplified if the data is "tidy" (Wickham, 2014). This tidiness begins with how the experimental assay is set up, for example by using a fixed pipetting scheme when working with 96-well microtiter plates. Once these steps have been taken, the quality of the experiment, the quality of the data and the quality of the model need to be independently assessed.

*How to report experimentally obtained high-quality data on an enzyme reaction*

With the successful completion of experiments, measured data need to be reported, either in an internal report, an open publication or a data repository. When reporting ERPs, adherence to the STRENDA guidelines (Beilstein Institute, no date) is recommended by a growing number of journals and repositories. Adherence to these guidelines is also useful for internal data reporting because it prevents omission of experimental parameters that are critical for repeatability of results (Halling *et al.*, 2018). To further ensure repeatability, it is also advisable to draft a standard operating procedure (SOP), for example by following the guidelines from Hollmann and co-workers (Hollmann *et al.*, 2020). SOPs are different from materials and methods sections because they directly provide step by step instructions for use in the laboratory and facilitate the interlinking of the obtained experimental data. This link between data and protocol is essential for interpreting and understanding results (Wolstencroft *et al.*, 2017).

Data volumes are quickly increasing to levels that can no longer be queried by humans, but require the use of machines, i.e. computers and algorithms. Therefore, efforts are ongoing to ensure data can be read by machines as well. To make data readable by humans and machines, data must be FAIR (findable, accessible, interoperable and reusable).(Wilkinson *et al.*, 2016) FAIR data is associated to FAIR metadata, such as experimental conditions, and FAIR vocabulary (Wilkinson *et al.*, 2016). These concepts provide context to the data so that machines can "understand" and work with the data. For example, to make a machine "understand" what ERPs mean, it needs to have access to metadata that uses an ontology[75] to describe the concept of an enzyme reaction, the context of such an experiment, such as experimental conditions, and the obtained data. Experimental conditions can be collected using electronic lab note books (ELNs) or stored in FAIR data repositories by linking SOPs to the raw data. FAIR reporting of data and metadata does not automatically mean that the data is "open," because data in a FAIR repository may be made accessible only to a select group of users.

"Open" is a term frequently used together with the terms "science",(Fecher and Friesike, 2014) "data" (Murray-Rust, 2008), "access"(Schiltz, 2018) and "source" (Opensource.com, no date; Delano, 2005). Open source is historically the oldest concept and stems from the concept of "source code". Source code is a series of human readable instructions that can be used to program a computer. Open source nowadays refers to something, typically software or code, anyone can modify and share because its design is publicly accessible.(Opensource.com, no date) However, unlike FAIR data, open data does not have to be structured in a specific manner or be in any way reusable or readable by humans and machines. The marriage of open and FAIR should ensure that the quality of data can be better assessed. Additionally, open and FAIR data on ERPs should be re-analyzable to confirm the obtained results, meaning that the modelling steps should be "re-editable code" (Hinsen, 2018). In this spirit follows an open, FAIR and re-editable example of the kinetic modelling of an enzyme reaction.

*An open, FAIR and re-editable example of the kinetic modelling of an enzyme reaction*

The reaction of the NADPH-dependent ketoreductase Gre2p (Genes de respuesta a estres, EC 1.1.1.283, sequence ID AJT71311.1) can be used in the asymmetric synthesis of chiral alcohols with excellent enantioselectivities (Müller *et al.*, 2010; Bitterwolf *et al.*, 2019). The conversion of 5-nitrononane-2,8-dione (NDK) to the preferentially produced (5S,8S)-anti hydroxyketone (HK, **Figure 4A**) has recently been characterized in detail using UV-Vis spectrophotometry and ITC (Ott *et al.*, 2021). This reaction is used as an example to illustrate some of the challenges discussed above. Initially, all analysis steps were performed manually using OriginPro (2020b (64-bit) 9.7.5.184, Academic) (Gygli and Ott, 2021a). To ensure the repeatability, replicability and reproducibility of the analysis steps and eliminate manual steps while also providing documentation of the entire modeling process, a Jupyter Notebook was created (https://fairdomhub.org/investigations/483).

A $K_m$ value around 1 mM has previously been reported for the conversion of NDK by Gre2p (Burgahn *et al.*, 2020). Therefore, [$S_0$] concentrations of 50 µM – 50 mM NDK were chosen to perform the experiment, with an enzyme concentration of 25 nM (Gygli and Ott, 2021c, 2021b). Each reaction was performed in four technical replicates and observed for almost 2h (**Figure 4B**). The artefacts present at the beginning of the measurements (up to 400 s) necessitated the exclusion of these datapoints from the analysis. Also, the reaction rate visibly slowed down after 750 s, indicating that initial rate conditions ([$S_0$]>>[P] and [$S_0$]>>[$E_0$]) were no longer true. Therefore, this data was not included in the analysis, leaving 11 datapoints that could be used for the linear fits to calculate initial rates (**Figure 4B**, inset). Quality of the linear fits was assessed using $r^2$, and a cutoff of 0.8 was used. This cutoff meant that data for [$S_0$] = 0.5 mM, [$S_0$] = 0.1 mM and [$S_0$] = 1.0 mM of replica 2, 3 and 4, was removed before calculating the average rates and standard deviations (**Figure 4C**). These rates were then plotted in a Michaelis-Menten plot and fitted with the Michaelis-Menten-equation (**Figure 4D**). A simple error propagation considering how the standard deviation of the linear fit affects the Michaelis-Menten-fit was included. In this error propagation, the distinction was made between the 1-σ confidence interval and the standard deviation of the fit. The 1-σ confidence interval estimates how likely it is that the "true" parameters are found (yellow in **Figure 4D**). The standard deviation of the fit only estimates how accurately the Michaelis-Menten equation represents the data (grey in **Figure 4D**). With this Jupyter Notebook, the impact of potentially low quality datapoints, as judged for example by the low $r^2$ values of the linear rate fits, can easily be established. Such low quality datapoints can then be removed to improve the kinetic model. It is also possible to change the "fitting window" of the linear fit, while documenting all these steps clearly (**Figure 5A**). The Jupyter Notebook is made available in a FAIR repository (https://fairdomhub.org/investigations/483) and in the SI to this review.

Note that the concentration of NADPH, 300 µM, used in this experiment is far below the $K_m$ value (2.4 mM), indicating that Gre2p is not saturated with NADPH under these conditions. Therefore, ITC has been used for additional kinetic measurements and to study the binding processes involved in the reaction. From these experiments, it has been found that Gre2p uses an ordered, sequential mechanism and that the enzyme suffers from substrate inhibition or product inhibition, depending on the composition of the reaction buffer. Also, DLS has been used to control the quality of the enzyme sample, and a Selwyn test has been performed to determine that no enzyme inactivation occurred (Ott *et al.*, 2021).

### *What to do with high quality data?*

Data becomes valuable if it is compared and combined to gain additional insight. For example, it is likely that increased data volumes and exchange of data have enabled the detection of the reproducibility crisis (Baker, 2016). Overcoming the reproducibility crisis of ERPs should be possible with the measures discussed above. So, what can large volumes of FAIR and open data be used for, and how can they be compared and combined to gain additional insight into enzyme reactions? The widespread use of ML artificial intelligence (AI) or deep learning (DL), implies that these algorithms can be used for anything and anywhere, from space traffic (European Space Agency, no date; Uriot *et al.*, 2021) to the email traffic on earth (Dada *et al.*, 2019) to enzyme engineering (Mazurenko, Prokop and Damborsky, 2020; Siedhoff, Schwaneberg and Davari, 2020).

The terms ML, AI or DL are often used interchangeably, yet there exist considerable differences between them. Historically, AI has been defined by Alan Turing in the "imitation game" in 1950, as a machine that can interact with a human without the human realizing that they are interacting with a machine (Turing, 1950). ML, an equally old concept, is a part of AI that has a much narrower focus and is used to solve very well-defined problems that follow a clear set of rules, such as checkers or chess. DL is a subfield of ML that studies neural networks. A form of ML frequently used is "supervised ML" (sML), whose goal it is to predict unknown properties ("labels") of data based on a set of known properties and associated "features". Labels are the target or output variable that the model should predict. For proteins and enzymes, examples of such labels are thermostability or solubility (Yang, Wu and Arnold, 2019). Features are the measurable properties that describe an object and have a link to the label of interest. For example, the features "unfolding free energy change" and "melting temperature change" have been used to predict the label "thermostability of proteins" (Jia, Yarlagadda and Reed, 2015).

The typical usage of sML involves five steps. (I) Define a workable research question to train the sML model, (II) use high-quality data able to answer the research question, (III) train the model using a representative "training set" of the data to build a model, (IV) predict the desired property for the unused data ("test data") using the model, and finally, (V) apply the model to completely new data for which the desired property is unknown. In this fifth step, the quality and power of ML models, but also the problems they face become evident. An ML model is nothing else than a mathematical equation that can be used to predict properties of data. If the data used to create the model contains biases, these biases are also "learned" (Cirillo et al., 2020). Therefore, care must be taken to balance the data and also include negative results. The sML model can only predict the label(s) it was trained for, and the data it was trained on may be too narrow for a broader application of the ML model. For example, models trained on one enzyme family may not accurately predict properties of another enzyme family. Most importantly, ML algorithms often operate as "black boxes", meaning that the reasons for the selection of a specific model often remain a mystery. This can lead to limitations in the application of these algorithms (Castelvecchi, 2016; Doshi-Velez and Kim, 2017; Poon and Sung, 2021). Bearing these limitations in mind, it is therefore crucial to not expect ML to solve problems unsolvable by humans, but to generate data with a specific question in mind, and ensure the question is suitable to be answered by ML.

Most ML algorithms are "data hungry" and typically require thousands of datapoints to build models that make accurate predictions. Such large data volumes of sufficient quality can be difficult and time consuming to obtain manually, therefore automation is needed. Automation of enzyme kinetics has previously been described (Lewis, Tallman and McGuinness, 2001; Bonowski *et al.*, 2010; Klimeš *et al.*, 2017), but appears to still be underused. This may be in part due to the complexity of the data and the increased data volumes, including the requirement for automated data analysis and management.

To pave the way for a rational understanding of enzyme reactions, enzyme mechanisms and enzyme reaction engineering (Sudar and Blažević, 2021), manual and repetitive steps in experiments and data analysis should be replaced by automation and ERPs should be reported in a standardized manner. These measures should enable scientists to more systematically study the effects of complex interactions of enzymes with their reaction components (Scheper *et al.*, 2021).

*Conclusion*

High-quality, reproducible, large and diverse data volumes are needed to rationalize the complexity of enzyme reactions and their dependence on reaction conditions. Reproducible and automated measurement and kinetic modelling of this data are needed to enable the collection of such large data volumes.

*Supplementary Information:*

**Jupyter Notebook** to Design an initial rate experiment (Design_an_Initial_Rate_Experiment.ipynb and Design_an_Initial_Rate_Experiment.pdf)

**Jupyter Notebook** to analyze an initial rate experiment (MichaelisMentenNotebook.ipynb and MichaelisMentenNotebook.pdf)

**Raw data** needed by the Jupyter Notebook to analyze an initial rate experiment (MM_25nM_Gre2p_NDK_NADPH_absorbance_340nm_cleaned.csv) This file needs to be placed in a folder "data" in the same location as the Jupyter Notebook.

**Functions** needed by the Jupyter Notebook to analyze an initial rate experiment (MMKinetics.py) This file needs to be placed in a folder "classes" in the same location as the Jupyter Notebook.

All supporting files are also available on FAIRDOMHub; [https://fairdomhub.org/investigations/483](https://fairdomhub.org/investigations/483)).

*Funding*

This work was funded by the Federal Ministry of Education and Research (BMBF) and the Baden-Württemberg Ministry of Science as part of the Excellence Strategy of the German Federal and State Governments. This work was also supported through the Helmholtz program "Materials Systems Engineering" under the topic "Adaptive and Bioinstructive Materials Systems".

*Notes*

The author declares no competing financial interest.

*Acknowledgements*

The author thanks Nico Henkenhaf for his contribution to the Jupyter Notebook to analyze an initial rate experiment.

doi:10.3389/fbioe.2015.00161.

Lorber, B. *et al.* (2012) 'Protein analysis by dynamic light scattering: Methods and techniques for students', *Biochemistry and Molecular Biology Education*, 40(6), pp. 372–382. doi:https://doi.org/10.1002/bmb.20644.

Lorsch, J.R. (2014) 'Practical steady-state enzyme kinetics', in *Methods in Enzymology*. 1st edn. Elsevier Inc., pp. 3–15. doi:10.1016/B978-0-12-420070-8.00001-5.

Ma, B. and Nussinov, R. (2013) 'Structured Crowding and Its Effects on Enzyme Catalysis', in Klinman, J. and Hammes- Schiffer, S. (eds) *Dynamics in Enzyme Catalysis*. Berlin, Heidelberg: Springer Berlin Heidelberg, pp. 123–137. doi:10.1007/128_2012_316.

Ma, S.K. *et al.* (2010) 'A green-by-design biocatalytic process for atorvastatin intermediate', *Green Chemistry*, 12(1), pp. 81–86. doi:10.1039/B919115C.

Mazurenko, S., Prokop, Z. and Damborsky, J. (2020) 'Machine Learning in Enzyme Engineering', *ACS Catalysis*, 10(2), pp. 1210–1223. doi:10.1021/acscatal.9b04321.

McArthur, S.L. (2019) 'Repeatability, Reproducibility, and Replicability: Tackling the 3R challenge in biointerface science and engineering', *Biointerphases*, 14(2), p. 020201. doi:10.1116/1.5093621.

McDonald, A.R. *et al.* (2022) 'Undergraduate structural biology education: A shift from users to developers of computation and simulation tools', *Current Opinion in Structural Biology*, 72, pp. 39–45. doi:10.1016/j.sbi.2021.07.012.

Müller, M. *et al.* (2010) 'Highly efficient and stereoselective biosynthesis of (2S,5S)-hexanediol with a dehydrogenase from Saccharomyces cerevisiae', *Organic & Biomolecular Chemistry*, 8(7), pp. 1540–1550. doi:10.1039/B920869K.

Murphy, E.F., Gilmour, S.G. and Crabbe, M.J.C. (2002) 'Effective experimental design: Enzyme kinetics in the bioinformatics era', *Drug Discovery Today*, 7(20), pp. 187–191. doi:10.1016/S1359-6446(02)02384-X.

Murray-Rust, P. (2008) 'Open data in science', *Serials Review*, 34(1), pp. 52–64. doi:10.1080/00987913.2008.10765152.

Olsen, H.S. and Falholt, P. (1998) 'The role of enzymes in modern detergency', *Journal of

*Surfactants and Detergents*, 1(4), pp. 555–567. doi:10.1007/s11743-998-0058-7.

Opensource.com (no date) *What is open source?* Available at: https://opensource.com/resources/what-open-source (Accessed: 15 October 2021).

Ott, F. *et al.* (2021) 'Toward Reproducible Enzyme Modeling with Isothermal Titration Calorimetry', *ACS Catalysis*, 11(17), pp. 10695–10704. doi:10.1021/acscatal.1c02076.

Plant, A.L. *et al.* (2014) 'Improved reproducibility by assuring confidence in measurements in biomedical research', *Nature Methods*, 11(9), pp. 895–898. doi:10.1038/nmeth.3076.

Poggi, C.G. and Slade, K.M. (2015) 'Macromolecular crowding and the steady-state kinetics of malate dehydrogenase', *Biochemistry*, 54(2), pp. 260–267. doi:10.1021/bi5011255.

Poon, A.I.F. and Sung, J.J.Y. (2021) 'Opening the black box of AI-Medicine', *Journal of Gastroenterology and Hepatology (Australia)*, 36(3), pp. 581–584. doi:10.1111/jgh.15384.

Robinson, P.K. (2015) 'Enzymes : principles and biotechnological applications', pp. 1–41. doi:10.1042/BSE0590001.

Romero, E. *et al.* (2018) 'Same Substrate, Many Reactions: Oxygen Activation in Flavoenzymes', *Chemical Reviews*, 118(4), pp. 1742–1769. doi:10.1021/acs.chemrev.7b00650.

Roy, S. *et al.* (2020) 'Discovery of Harmaline as a Potent Inhibitor of Sphingosine Kinase-1: A Chemopreventive Role in Lung Cancer', *ACS Omega*, 5(34), pp. 21550–21560. doi:10.1021/acsomega.0c02165.

Rungsrisuriyachai, K. and Gadda, G. (2009) 'A pH switch affects the steady-state kinetic mechanism of pyranose 2-oxidase from Trametes ochracea', *Archives of Biochemistry and Biophysics*, 483(1), pp. 10–15. doi:10.1016/j.abb.2008.12.018.

Scheper, T. *et al.* (2021) 'Digitalization and Bioprocessing: Promises and Challenges BT - Digital Twins: Tools and Concepts for Smart Biomanufacturing', in Herwig, C., Pörtner, R., and Möller, J. (eds). Cham: Springer International Publishing, pp. 57–69. doi:10.1007/10_2020_139.

van Schie, M.M.C.H. *et al.* (2021) 'Applied biocatalysis beyond just buffers - From aqueous to unconventional media. Options and guidelines', *Green Chemistry*, 23(9), pp. 3191–3206.
23

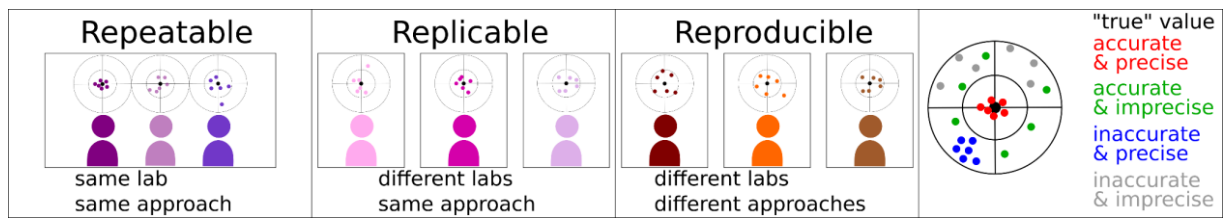

*Figure 1: Data quality and the distinction between repeatability, replicability and reproducibility of accurate and precise measurements to determine a "true" value. Note that precise measurements are not a guarantee that the "true" value is found. The "trueness" of a measurement, and therefore its accuracy can only be established through reproducible measurements.*

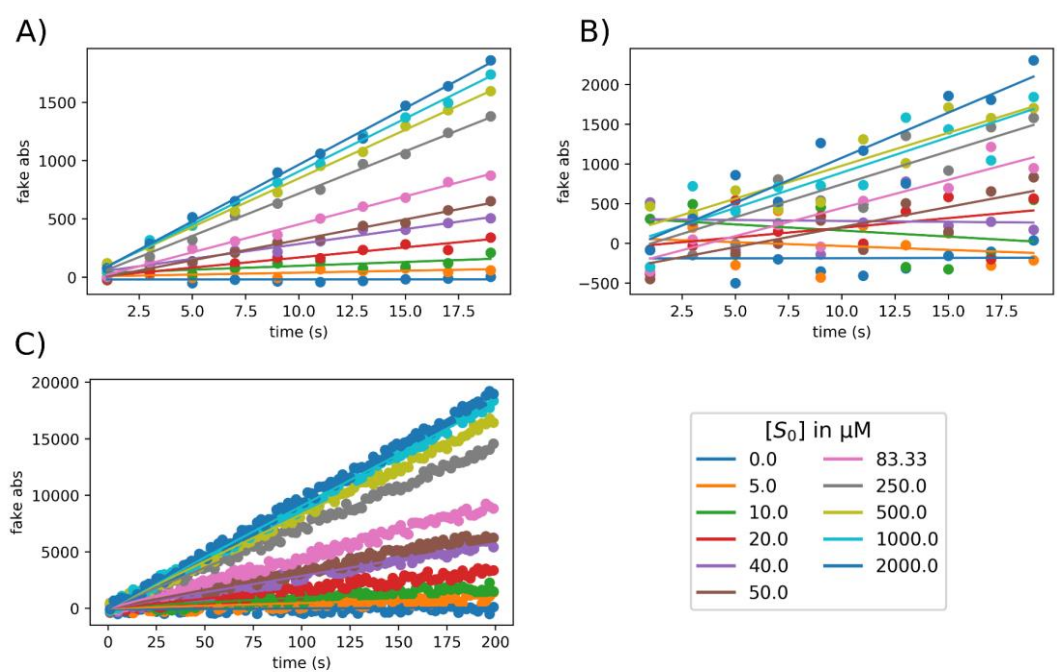

*Figure 2: Illustration of the impact of the number of datapoints and the noise in the measurement on the linear fit. A) Ten datapoints are sufficient to obtain a reasonable, linear fit if there is little noise in the measurements. B) If there is considerable noise in the measurement, 10 datapoints are no longer sufficient to obtain a reasonable, linear fit. C) Increasing the number of datapoints improves the linear fit of the data even if there is considerable noise. Data was simulated using the Jupyter Notebook to design an initial rate experiment (https://fairdomhub.org/investigations/483). [$S_0$] are the same as in **Figure 3D**.*

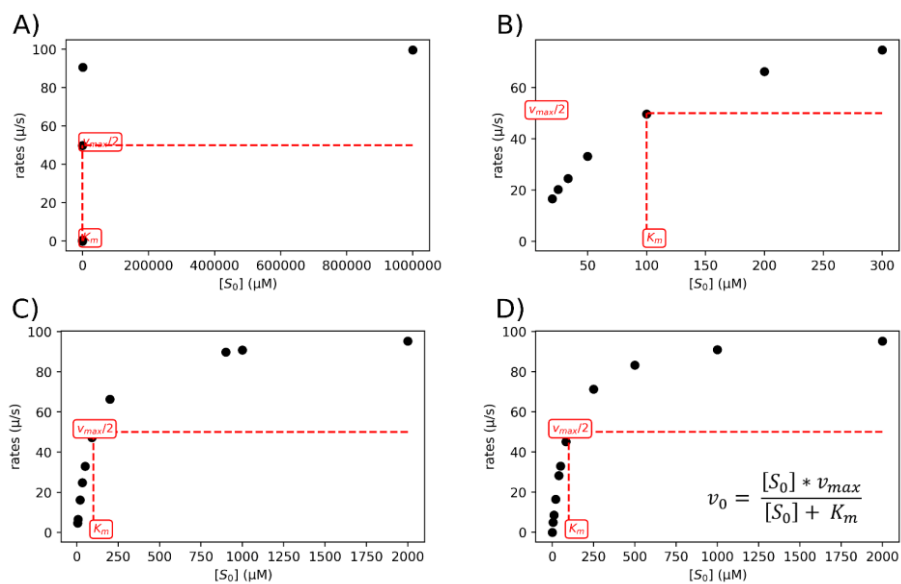

*Figure 3: How to design an initial rate experiment to obtain high-quality data. Data was simulated using $K_m$ = 100 µM and $v_{max}$ = 100 µM /s. A) Zero-round experiment with 5 widely spaced [$S_0$] (0, 0.01, 1, 100, 1000 and 1000000 µM). B) First-round experiment with badly chosen [$S_0$] ($K_m$/5, $K_m$/4, $K_m$/3, $K_m$/2, $K_m$, $K_m$*2 and $K_m$*3). C) First-round experiment with well-chosen [$S_0$] ($K_m$/20, $K_m$/15, $K_m$/5, $K_m$/3, $K_m$/2, $K_m$/1.1, $K_m$*2, $K_m$*9, $K_m$*10 and $K_m$*20). D) Gold-round experiment with ideally chosen [$S_0$] (0 µM, $K_m$/20, $K_m$/10, $K_m$/5, $K_m$/2.5, $K_m$/2, $K_m$/1.2, $K_m$*2.5, $K_m$*5, $K_m$*10 and $K_m$*20). Data was simulated using the Jupyter Notebook to analyze an initial rate experiment (https://fairdomhub.org/investigations/483).*

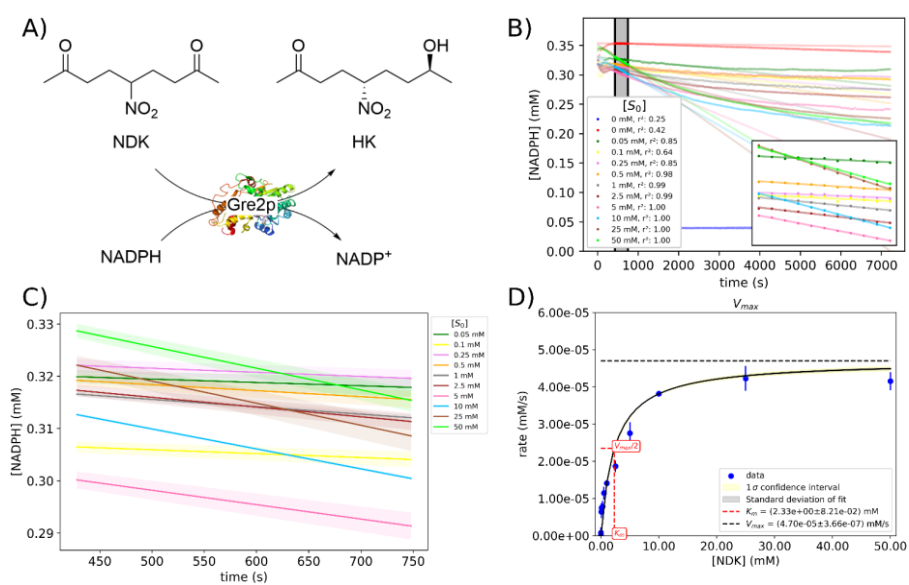

*Figure 4: An open, FAIR and re-editable example of the kinetic modelling of an enzyme reaction. A) Gre2p-catalyzed conversion of 5-nitrononane-2,8-dione (NDK) to the preferentially produced (5S,8S)-anti hydroxyketone (HK). B) Illustrative replica for an initial*

*rate experiment. Due to the noise in the beginning and the end of the measurement, the linear fit is only performed on a selection of the data, which is shaded in grey in the main plot and shown for clarity in the inset. C) Linear fits including standard deviation of the four replicas. D) Michaelis-Menten plot of the initial rates at different [$S_0$] to yield $K_m$ and $v_{max}$. This plot includes the standard deviation of the individual rates (blue bars), the standard deviation of the fit (grey shading) and the confidence interval of the data (yellow shading). See text for details.*